%% file: DASH EuroVis 2022/mainbody-teaser.tex
\title[\toolname: Visual Analytics for Debiasing Image Classification]%
      {\toolname: Visual Analytics for Debiasing Image Classification\\ via User-Driven Synthetic Data Augmentation}

\author[B. C. Kwon et al.]
{\parbox{\textwidth}{\centering Bum Chul Kwon$^{1}$\orcid{0000-0002-9391-6274}, Jungsoo Lee$^{2}$, Chaeyeon Chung$^{2}$, Nyoungwoo Lee$^{2}$\orcid{0000-0001-5660-133X}, Ho-Jin Choi$^{2}$\orcid{0000-0002-3398-9543}, Jaegul Choo$^{2}$\orcid{0000-0003-1071-4835}
        }
        \\
{\parbox{\textwidth}{\centering $^1$IBM Research, Cambridge, Massachusetts, United States\\
         $^2$ KAIST, Daejeon, Republic of Korea
       }
}
}

\begin{document}

\teaser{
\fbox{
 \includegraphics[width=\linewidth]{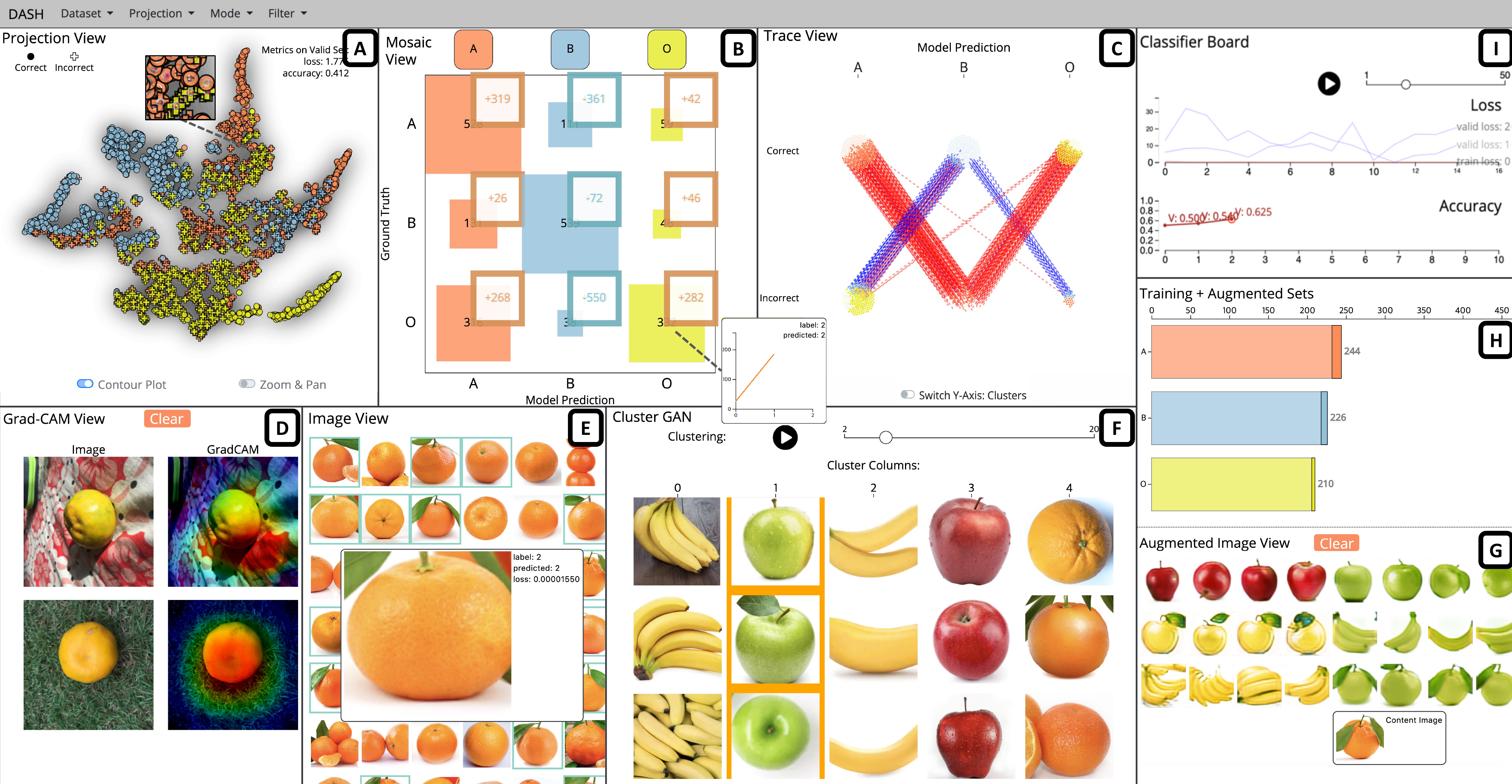}
}
 \centering
  \caption{An overview of \toolname: (A) \proj shows the latent space representation of images using t-SNE; (B) \mosaic summarizes the performance differences between two previously trained classifiers; (C) \trace shows how the two classifiers predict individual images differently (red: incorrect to correct, blue: correct to incorrect); (D) \gradcam shows the feature importance of images as heatmaps; (E) \imglist shows a list of selected images; (F) \cluster shows clustering results which can be used to translate visual features of images to other images using \xgan; (G) \augmentedlist shows newly created images for retraining and (H) shows the summary of the new images; (I) \class shows the performance of different classifiers.}
\label{fig:teaser}
}

\maketitle
\begin{abstract}
   Image classification models often learn to predict a class based on irrelevant co-occurrences between input features and an output class in training data. We call the unwanted correlations ``data biases,'' and the visual features causing data biases ``bias factors.'' It is challenging to identify and mitigate biases automatically without human intervention. Therefore, we conducted a design study to find a human-in-the-loop solution. First, we identified user tasks that capture the bias mitigation process for image classification models with three experts. Then, to support the tasks, we developed a visual analytics system called \toolname that allows users to visually identify bias factors, to iteratively generate synthetic images using a state-of-the-art image-to-image translation model, and to supervise the model training process for improving the classification accuracy. Our quantitative evaluation and qualitative study with ten participants demonstrate the usefulness of \toolname and provide lessons for future work.


\end{abstract}

\input{sections/01-intro}
\input{sections/03-task}
\input{sections/04-system}

\section{User Study}
 
We conducted a user study, where participants perform bias mitigation tasks on two datasets, 1) fruit dataset~\cite{fruit-dataset1, fruit-dataset2} (450 images) and 2) cartoon dataset~\cite{cartoon-dataset} (680 images), and provide their insights about \toolname through interviews at the end.
We initially trained models with low accuracy less than 55\% with biases on colors of fruits and sunglasses of cartoon characters, respectively.
Participants were asked to identify the source of biases and mitigate them using \toolname.
We recruited ten participants (9 graduate students and a recent graduate; 6 males and 4 females; mean age of 24.5), who are studying/working in computer vision for at least 6 months (10.7 months of working experiences in average). They were instructed with a video and provided with a tool on a toy dataset o that they can learn how to use the tool.
Each participant took two hours to complete the study and received \$12.5 per hour for the reward.
All users performed their tasks with a Macbook Pro (16-inch, 2019) monitor with a screen resolution of 2560$\times$1600. 
We also used one GPU (NVIDIA TITAN Xp 12GB VRAM) for the computation and Chrome (v. 85.0.4183.102) for the browser.

All participants successfully achieved the test accuracy of 90\% for the cartoon dataset, and seven out of ten participants (P2, P4, P6-10) reached the test accuracy of 65\% for the fruit dataset. 
In the process, participants retrained 2.1 and 3.3 times on average for cartoon and fruit datasets, respectively. 
Wilcoxon Signed Rank Test indicates that there was a statistically significant difference in retraining iterations between the cartoon dataset and the fruit dataset (\textit{p}$=$0.048).
Understandably, participants perceived the fruit dataset more difficult and did more poorly on it than the cartoon dataset because the fruit dataset includes real images of fruits.

Participants shared their experiences with \toolname.
Here we provide some areas for improvements based on their comments.
Three participants (P2, P4, P5) pointed out that \toolname requires domain expertise and prior experiences in deep learning and data augmentation. 
P2 added, ``Novice users would take more time to learn how to use \toolname. They may need hands-on tutorials for a longer period of time. The instructional video was useful to understand the tool.''
Participants also described why some views were difficult to use.
P6 reported ``While exploring the images, I observed only subtle differences in visual characteristics between clustering results with different $K$s.''
Additionally, we observed that prior knowledge about particular views in \toolname may prevent participants from using the views.
For example, P7 finished his tasks for both datasets without using \gradcam. He said ``I do not trust the robustness and usefulness of GradCAM~\cite{gradcam} because I didn't find the technique useful in the past. 
This prior knowledge prevented me from using \gradcam and I trusted my own judgment when I inspected individual images.''

Participants also shared future ideas to improve the bias mitigation processes using \toolname.
First, three participants (P1, P3, P4) wanted to keep the data augmentation history. 
While repeatedly generating and discarding images, the participants easily forgot what they already did or what they should do.
Thus, the participants wanted to keep track of their previous attempts in order to save time and efforts.
Three participants (P3-5) also reported that they wanted to separately analyze images which were frequently misclassified over previously trained models.
In that way, they can further investigate why the model keeps making mistakes and derive a strategy to mitigate the specific biases.

\section{Conclusions}
\label{sec:08-conclusions}

Our work studies a visual analytic approach to tackle the problem of debiasing image classification models through data augmentation. 
We designed \toolname by conducting a design study with experts in deep learning. 
\toolname integrates the state-of-the-art image translation technique with various views as a unified system which saves time and cognitive efforts of users. 
In particular, various views of \toolname lead users to gain key insights that are required for debiasing.
The user study and the quantitative evaluation demonstrate that \toolname can provide users with capabilities to effectively solve real-world biases in image data. 
Future work can investigate ways to help novice users learn how to use the tool.
Our experiment is limited because we used a small dataset with relatively simple biases due to time constraints.
Future work can also investigate the use of bias mitigation tools like \toolname on a large-scale dataset like ImageNet~\cite{imagenet} for a longer period of time through a long-term case study~\cite{shneiderman_strategies_2006}.
Future studies can embed such tools in notebook environments like Jupyter Notebook so that data scientists can develop their own models.
Users aim to achieve high quality for translated images, so it will be useful to develop a user interface to retrain \xgan interactively.
The study shows task analysis, tool design, and user experiments which can be useful to conduct future studies on developing visual analytics tools for bias mitigation in image classification models.

\bibliographystyle{eg-alpha-doi}  
\bibliography{egbibsample}        

\newpage

\end{document}

%% file: sections/01-intro.tex
\section{Introduction}

Image classification models often learn to predict an output class based on irrelevant features co-occurring with the class within images in training data~\cite{singh_dont_2020}.
We call the undesirable correlation between some visual features and class labels in training data as ``data biases,'' and refer to such visual features causing the biases as ``bias factors.''
For example, as illustrated in the previous literature~\cite{bahng2020learning}, many images of `frogs' in training data are taken with `swamps' in the background. 
Image classification models often make mistakes by predicting the class of frogs based on swamps in the background.
In this example, swamp is a bias factor that causes the image classification model to be biased for the class label frogs.
Though biased models may provide high accuracy in training data, they can result in fatal errors on unseen data beyond training data. 
Therefore, it is important for data scientists to identify and mitigate biases in models before deploying them.

To resolve data biases, image classification models need to unlearn irrelevant features and learn more important features that are related to the output class. 
In this context, data augmentation can be a viable solution that can generate synthetic images by artificially combining existing ones into new images. 
However, it is difficult to automatically identify bias factors of given models and generate images that can effectively target and remove the unwanted correlations.
The process is often iterative and labor-intensive because data scientists need to inspect the models to discover bias factors among many potential features, generate images by augmenting existing images, and re-train and evaluate the model so that it can achieve a higher performance in testing data.

In this paper, we conducted a design study with thirteen experts in image classification to develop a visual analytics system for the model debiasing problem.
First, we analyzed the user tasks with three experts to understand the model debiasing process.
Second, based on the user tasks, we developed a visual analytics system called \toolname (\textbf{D}ata \textbf{A}ugmentation \textbf{S}ystem for \textbf{H}uman-in-the-loop).
\toolname allows data scientists to visually identify bias factors among non-trivial visual attributes of images (e.g., colors and object presence). 
Using \toolname, they can also synthesize new images by translating target attributes using a state-of-the-art image-to-image translation technique called \xgan~\cite{exploregan} and evaluate the quality of the generated images.
Finally, \toolname allows them to retrain the model with the newly created images and evaluate the performance of the revised model against previous models.
To evaluate \toolname, we conducted a user study with ten machine learning experts on two real-world datasets.
The results demonstrate that \toolname helps data scientists discover and mitigate biases of image classification models. 

Our main contributions include:
\begin{itemize}
    \item We identify user tasks with three experts that capture the user-driven debiasing process for image classification models.
    \item We present a visual analytics system called \toolname that allows users to identify the bias factors, to synthesize new images using image-to-image translation, and to visually supervise the model retraining process.
    \item We conduct a qualitative evaluation with ten machine experts to show the effectiveness of using \toolname for debiasing image classification models.
\end{itemize}

\begin{figure}[t!]
        \includegraphics[width=\linewidth, clip]{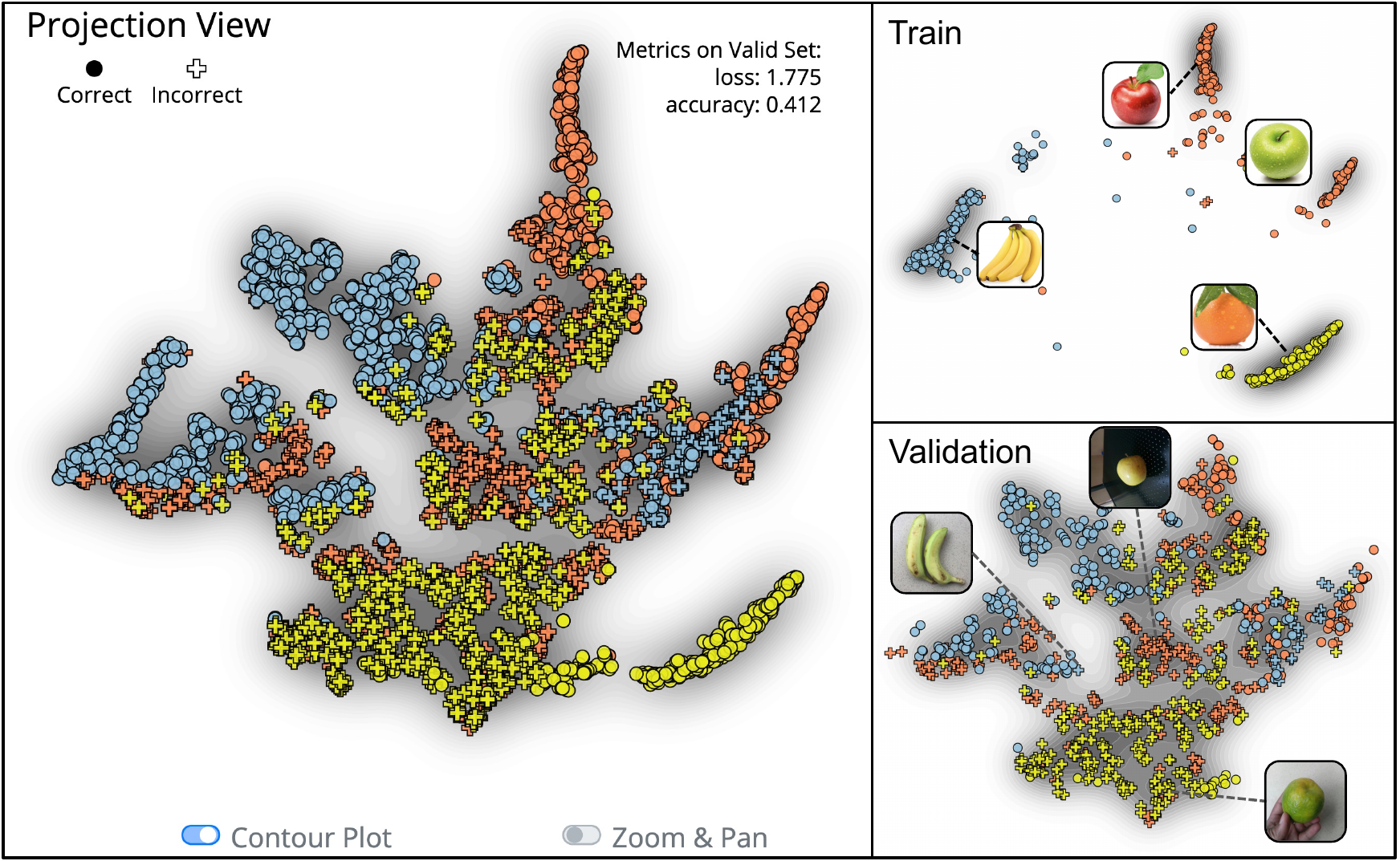}
    \caption{In \proj, users find clusters of images. Circles and crosses represent correctly classified and misclassified images, respectively. The color inside each point shows its true class label, and the color of the stroke (border) shows its predicted class. Users can zoom in to view actual images.}
    \vspace{-.5cm}
    \label{fig:projection}
\end{figure}

%% file: sections/03-task.tex
\section{User Tasks: Model Debiasing Process}
\label{sec:03-taskanalysis}

We derived the tasks based on discussions among the co-authors of this work, who are experts in the fields of computer vision.
The following tasks represent high-level objects that users need to perform in order to mitigate biases in image classification models.
We assume that users already have trained an image classification model with a training dataset.

T1: \textbf{Discover data biases in training data.} 
With the initial model, users generate the accuracy of the model on test (unseen) data.
The model generates lower accuracy, so the users investigate the source of errors.
The errors usually originate from homogeneous distributions of training data, which include unintended correlation between visual attributes of images (swamp) and classes (frog) in training data.
Users filter data by the specific class labels that cause errors in test data and then derive irregularities.
They often use activation maps like GradCAM~\cite{gradcam} to highlight important regions of images that the model uses for the classification task.
After iterative exploration, they hypothesize the unwanted correlation between some visual features and class labels. 

T2: \textbf{Mitigate bias through data augmentation.}
Once users identified the sources of errors (target visual features to unlearn), users need to generate new images.
Users need to generate images of frogs with diverse backgrounds like street, house, and tree.
Users can use an image translation technique to change an attribute (swamp) of an image to another attribute (street) from other images without altering other attributes (frog).
Image translation models require a ``source'' image including the class label and a ``style'' image containing diverse attributes (e.g., street, house, tree) to be fused into the source image.
After training the image translation models, users evaluate how realistic the resulting image is. 
Once satisfied with the quality, users generate new images with diverse backgrounds and label them with the target class for retraining.

T3: \textbf{Retrain, evaluate, and steer the classifier.}
With the newly generated images, users retrain the image classification model.
In this process, users adjust model parameters and hyperparameters (e.g., epoch number, batch size, learning rate) to maximize the learning outcome.
After retraining the classifier, users evaluate the performance of classification model.
In addition to accuracy, users need to assess whether the revised model correctly classify images of the target label.
In particular, it is important to test whether the fused visual feature was helpful to resolve the target biases in the model.
This process cannot be done at once.
Some images that users generated may not help the model to improve their accuracy.
In the worst case, some images may decrease the accuracy of the model.
Then, users can discard the model instance and retrain the previous version of the model with newly generated images.

%% file: sections/04-system.tex
\section{\toolname: Visual Analytics for Data Augmentation}
\label{sec:04-system}

Based on the user tasks, we developed \toolname which includes multiple, coordinated views.
The following sections describe how the system supports the bias mitigation tasks. 

\subsection{\proj} 
\proj presents the overall distribution of the latent space in the training and validation sets with a two-dimensional scatter plot where each data point indicates each image.
Following the recent studies~\cite{ganviz, Choi:2019:AILA, embedding-projector, Chan:2018:CUDA, kwon-cluster}, we created a two-dimensional scatter plot of images using t-SNE~\cite{t-sne}. 
We used each image's latent space representation extracted from the last convolutional layer of the image classification model. 
By doing so, the representation of each image captures its high-level semantics (e.g., background colors, objects, texture)~\cite{netdissect}.
\autoref{fig:projection}~(a) shows that \proj separates images by the fruit types in the training set based on colors.

By exploring \proj, users can discover a noticeable difference in the data distributions between the training and validation sets (\textbf{T1}).
A contour plot of \proj indicates the estimated density of image clouds. 
In addition, users can check whether each data point is predicted correctly or incorrectly with the marker shape, a circle (correct) or a cross (incorrect) respectively, which are colored differently according to its ground truth class (\textbf{T1}). 
We allow users to zoom into a cluster, where each point turns into the actual image at the maximum zoom level. 
They can also gain the additional information of an image, such as its class label, predicted label, and prediction loss value, in a popup. 
They can also select a group of data points by using lasso-selection to load the actual images on \imgview right below \proj.
The validation set of \autoref{fig:projection}~(a) shows misclassified items, which include images of fruits in their unusual colors (e.g., green bananas).

\subsection{\gradcam}
\begin{figure}[t!]
   \centering
    \includegraphics[width=\linewidth, clip]{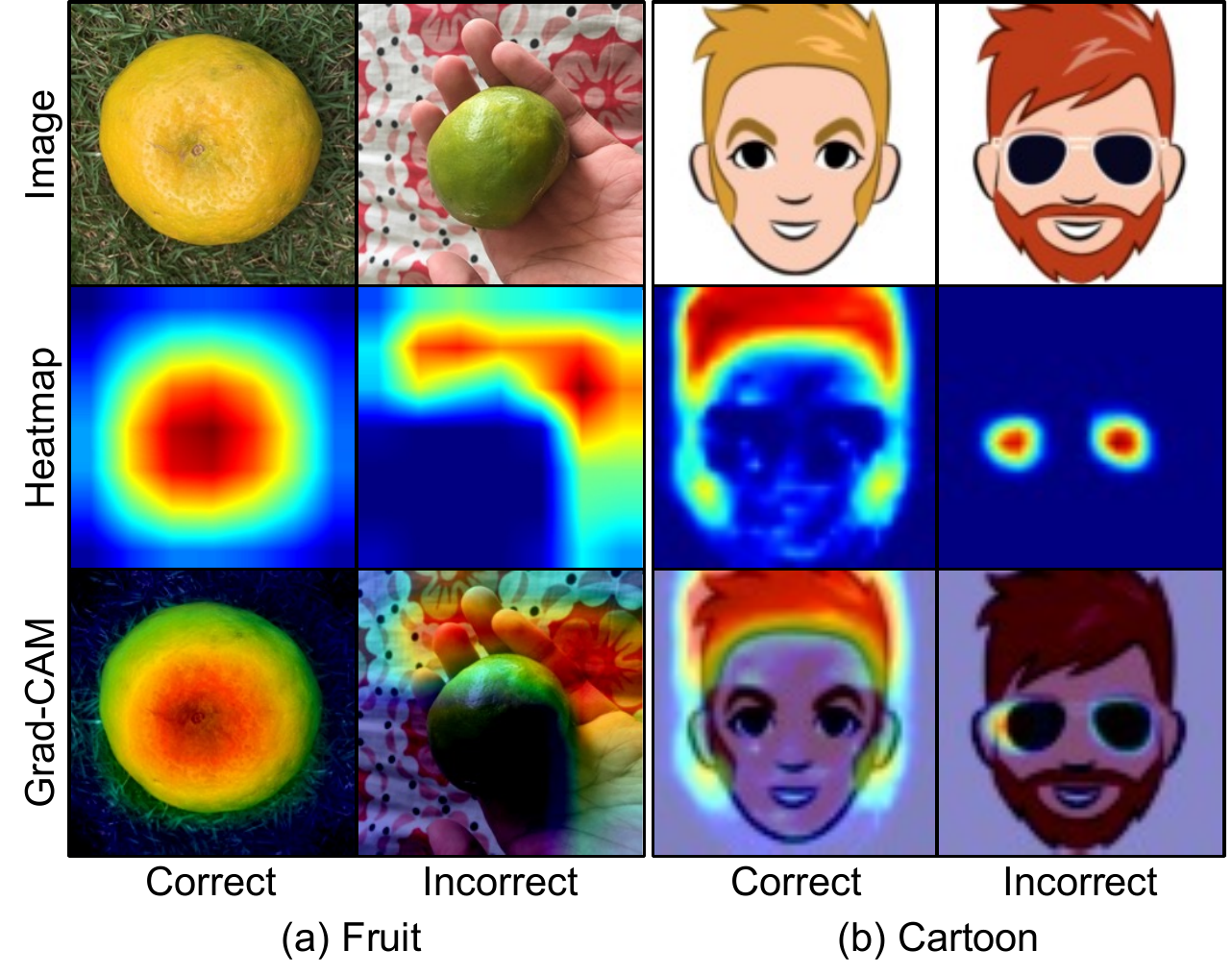}
    \caption{\gradcam shows the areas that the model gives the most attention as heatmaps (Red: High; Blue: Low). The three rows indicate i) original image; ii) heatmap; iii) original $+$ heatmap. Users can infer where the model should focus on by comparing the correctly classified and misclassified images in \gradcam.}
    \vspace{-.6cm}
    \label{fig:gradcam}
\end{figure}

\gradcam helps users to understand to which areas of an image the model attributes more importance while making decisions (\textbf{T1}). 
One can also consider using other existing explanation methods, such as saliency maps, Guided BackProp~\cite{guided_back_prop}, and Guided \gradcam~\cite{guided_gradcam}.
As \autoref{fig:teaser}~(D) shows, \gradcam shows the heatmap.
By interpreting Grad-CAM heatmaps over the images, users can estimate the regions that are correlated with its predicted class label.
For example, as the first column of \autoref{fig:gradcam}~(a) shows, the model accurately classifies an image of an orange with the focus on the orange. 
On the other hand, the model misclassifies another image of an orange as an apple in the second column of \autoref{fig:gradcam}~(a). 
\gradcam shows that the model focuses on the peripheral region of the image instead of the green orange in the center.

\subsection{\cluster}
\cluster presents groups of images by discovering clusters of the latent space vectors (\textbf{T2}). 
We chose \xgan~\cite{exploregan} over other image translation models (e.g.,~\cite{stargan,MUNIT,drit, pix2pix,cyclegan,stgan}) because the model can generate new images without predefined labels for style features (e.g., swamp).
\xgan clusters images into $K$ groups using K-Means clustering~\cite{k-means}, then uses the cluster indices as pseudo labels for style features shared by the images within each cluster. 
Then, \xgan can transfer visual features present in the images of a cluster to other images.

As \autoref{fig:teaser}~(F) presents, users can run clustering by adjusting the target number of clusters ranging from 2 to 20.
Once the clustering completes, it shows a table with columns (clusters) of $N$ representative images that are the closest to the centroid of their respective clusters.
Users choose a column (highlighted in orange in \autoref{fig:teaser}~(F)) to use the images in the cluster as a target style images and choose an image from \imgview as a source image in \autoref{fig:teaser}~(E).
After generating new images, users can validate the quality of the generated images on \augmentedlist as \autoref{fig:teaser}~(G) shows (\textbf{T2}).

\subsection{\class, \mosaic, and \trace}
\label{sec:mosaic}

\class allows users to visually supervise the retraining process and to navigate the results (\textbf{T3}).
While retraining, \class shows loss values of training and validation sets for every epoch in a line chart (red: training, blue: validation).
Moreover, \class enables users to switch back and forth among the previously trained models (\textbf{T3}).
By doing so, users can discard unsuccessful retraining attempts if necessary.

\mosaic shows the differences in classification results between two different models (\textbf{T3}).
Inspired by prior studies on confusion matrix~\cite{kapoor_interactive_2010, torkildson_visualizing_2013, alsallakh_visual_2014, ren_squares:_2017}, \mosaic highlights the cell-level differences using a confusion matrix.
Each cell in \mosaic changes its size in proportion to the number of images in the corresponding cell. 
This allows users to understand the overall model performance at a glance (\textbf{T1}).
Users can click on cells of interest by using CTRL $+$ Click, and it filters other views by the images in the cells. 
For instance, users can select the two cells in (2,1) and (2,3) of \autoref{fig:teaser}~(B) to inspect the images of `banana', which are misclassified as `apple' and `orange', respectively.

\trace summarizes how individual images are predicted differently from different model instances.
In \trace, points in the upper row are correctly classified images while those in the lower row are incorrect ones, as \autoref{fig:teaser}~(C) presents.
Red lines indicate that the items were previously predicted incorrectly; blue lines show that the items were previously predicted correctly.
By browsing across multiple models from different iterations of training, users can gain insights about 1) the changes between different iterations and 2) edge cases, where models constantly make mistakes (\textbf{T3}).
The insights can lead users to select the image group of interest and to analyze them in more detail to understand why they are predicted differently during different iterations of retraining.